\documentclass[aps,prb,singlecolumn,showpacs]{revtex4}
\usepackage{epsfig}
\usepackage{amsfonts,amsmath,amssymb}
\usepackage{graphicx}
\begin{document}
\title{From Cooper-pairs to resonating bipolarons}
\author{Julius~Ranninger}
\affiliation{Centre de Recherches sur les Tr\`es Basses Temp\'eratures,\\associ\'e \`a l'Universit\'e Joseph Fourier, CNRS, BP~166, 38042 Grenoble cedex 9, France}
\maketitle

\section{Introduction}
By far the most decisive experimental results which led the way in 
acquiring  a theoretical understanding of superconductivity was (i)~the Meissner 
effect and (ii)~the isotope shift in the critical temperature $T_c$.

Fritz and Heinz London early on recognized \cite{London-35} that the Maxwell 
equations had to be supplemented by a relation between the magnetic field applied 
to a superconductor  and the thereby induced macroscopic circulating 
currents, in order to account for the expulsion of such an externally applied 
magnetic field from the inside of the  superconducting sample. The shielding 
supercurrents necessarily  being a macrosocopic feature, Fritz London conjectured  this 
to be a manifestation of the Bose Einstein condensation, according to which 
the charge carriers making up such supercurrents are condensed into a single 
quantum state---now generally refered to as  macroscopic coherent quantum state.  
Invoking the Heisenberg uncertainty principle, the charge carriers making up such 
spatially homogeneous  supercurrents moving on huge orbitals, suggested 
that any order defining such a superconducting state must be an order in momentum 
rather than real space of the charge carriers.  Since 
it could be expected that the electrons participating in the supercurrents 
came from only a very small region in {\bf k}-space around the
Fermi surface, this insight served as a hint of vital importance in the 
construction of the ultimate solution for the superconducting ground state wave 
function \cite{BCS-57}. 

The other such vital hint helping to solve the ridle of the 
superconductivity phenomenon came from the theoretical work by 
H.~Fr\"ohlich \cite{Frohlich-50,Bardeen-51}, who 
showed that, inspite of the heavy mass of the ions and the repulsive 
Coulomb interaction between the electrons, two electrons near the Fermi 
surface  will interact attractively with each other. He suggested that the 
difference in energy beween the superconducting and the normal state of 
a metal could result from such electron lattice interaction. The 
experiments \cite{isotope-50} on the isotope shift of the critical temperature 
of superconductors, carried out at the same time and totally indepedent on any 
knowledge of these theoretical results, strongly supported this suggestion. 

Given the insight coming from the Meissner effect and the isotope shift 
in $T_c$, the problem then was to put that knowledge on a firm theoretical ground. 
This was done in the seminal work by Bardeen, Cooper and Schrieffer \cite{BCS-57}.
Inspite of the success in explaining the isotope shift of $T_c$ as arising from an 
effective lattice modulated electron-electron 
interaction \cite{Frohlich-50,Bardeen-51}, this initial work 
was in an order of magnitude conflict with the experimentally observed energy 
difference between the superconducting and the normal state.
This difficulty was overcome in a study considering the interaction of just two 
electrons above an immobile Fermi sea \cite{Cooper-56} and clearly 
indicated the instability of the Fermi sea as a result of the electron-lattice 
interaction. The final remaining task then was to  repeat this study for the 
ensemble of electrons in the Fermi sea. This was achieved in  what is now known 
as  the BCS wavefunction \cite{BCS-57}, constructed in such a way that the 
electrons optimally exploit the electron-lattice coupling in their 
scattering processes,  allbeit in full respect of the Pauli principle.
The BCS theory relates the appearence of the superconducting 
state to the appearence of Cooper-pair formation. This is manifest in the 
developement of a finite amplitude of the order parameter, which shows  up in 
form of a gap in the single particle electron specrum. The role 
of the phase of the order parameter in this superconducting BCS  ground state, 
which is a prerequisit of any coherent macroscopic quantum state and necessary to 
assuring the Meissner effect \cite{Ranninger-Thirring-63}, is hidden in this  
early BCS formulation  and which, in a more transparent way, was subsequently 
provided by P. W.~Anderson \cite{Anderson-58}. 

The BCS theory for electron-lattice induced superconductivty applies to 
systems with weak electron lattice interaction, implying a small  energy gap over 
Fermi energy ratio $\Delta(0)/\varepsilon_F\ll 1$ and the adiabatic regime 
$k_B\theta_D/\varepsilon_F \ll 1$, with $\theta_D$ denoting the Debye frequency.
The question which has occupied the superconductivity community for some time 
is to establish what happens to such a BCS superconductor if those conditions 
are no longer satisfied. Intuitively one could expect \cite{Chakraverty-79} that upon increasing the 
electron-lattice interaction would  ultimately lead to a local polaron type 
instability of the lattice structure surrounding the charge carriers and hence 
give rise to localization. On the other hand, provided one is in an 
anti-adiabatic regime (with a characteristic local lattice vibrational mode 
energy much bigger than the bare electron hopping integral), one obtains a 
superfluid phase of local electron-pairs, the 
socalled Bipolaronic Superconductivity \cite{Alexandrov-Ranninger-81}.
The cross-over between the BCS regime either to a superfluid state of tightly 
bound electron pairs or to their localization is presently a field of great 
interest. In the following sections we shall review the issues involved in such 
a physics.

\section{From amplitude to phase fluctuation controlled superconductivity}\label{phaseflucsup}

Let us consider a quite general electron-lattice coupling Hamiltonian 
\begin{eqnarray}
H &=& H_{\rm el} + H_{\rm ph} + H_{\rm el,ph} \nonumber \\
&=& \sum_{{\bf k} \sigma} (\varepsilon_{\bf k}-\mu) 
c^+_{{\bf k}\sigma}c^{\phantom +}_{{\bf k}\sigma}+\sum_{\bf q}\omega_{\bf q}\left(a^+_{\bf q}a^{\phantom +}_{\bf q} + \frac{1}{2}\right)
\nonumber \\
&+& \sum_{{\bf q}{\bf k}\sigma}V_{\bf q}c^+_{{\bf k+q}\sigma}
c^{\phantom +}_{{\bf k}\sigma}(a^{\phantom +}_{\bf q} + a^+_{-{\bf q}})
\end{eqnarray}
which  permits us to describe the essential features of BCS as well as of 
Bipolaronic 
Superconductivity in the simplest possible way. $c^{+}_{{\bf k}\sigma}, 
c^{\phantom{+}}_{{\bf k}\sigma}$ denote bare electron creation and 
annihilation operators for Bloch states characterized by momenta ${\bf k}$ 
and spin $\sigma$ and having a bare electron spectrum given by 
$\varepsilon_{\bf k}$.  $a^{+}_{\bf q}, a^{\phantom{+}}_{\bf q}$ denote  
creation, respectively annihilation operators for phonons with wave 
vectors ${\bf q}$ and having bare phonon frequencies $\omega_{\bf q}$. 
Finally, $V_{\bf q}$ denotes an effective electron-phonon coupling, the 
effect of which is twofold: (i) to ``delocalize'' the Bloch 
states, i.e., the electron momentum being no longer a conserved quantity speads
the resulting excitations around the wavevectors which characterize the 
system in the absence of the electron-phonon coupling, 
(ii)~to induce lattice deformations in the neighborhood of the momentary positions 
of the electrons on the lattice. Effective Hamiltonians which capture these
features can be constructed by use of unitary transformations  
$\tilde{H} = e^{-S}He^{S}$ with
\begin{equation}
S = \sum_{{\bf i}\,{\bf j}\,l\,\sigma} c^+_{{\bf i}\sigma} 
c^{\phantom +}_{{\bf j}\sigma}(\pi_{(j-i,\;l-i)}P_{\bf l} + 
\chi_{(j-i,\;l-i)}X_{\bf l}) 
\end{equation}
where 
\begin{eqnarray}
X_{\bf i}&=&\sum_{\bf q} \sqrt{\left({\hbar \over 2 N M \omega_{\bf q}}\right)}
(a^{\phantom +}_{\bf q}+ a^{+}_{-{\bf q}})e^{i{\bf q}\cdot {\bf R}_{\bf i}} \nonumber \\
P_{\bf i}&=& i\sum_{\bf q} \sqrt{\left({\hbar  M \omega_{\bf q} \over 2 N } \right)}
(a^+_{\bf q}-a^{\phantom +}_{-{\bf q}})e^{-i{\bf q}\cdot {\bf R}_{\bf i}}
\end{eqnarray}
denote the operators describing electron-phonon coupling induced local lattice 
deformations and shifts in the electron momenta respectively.

Quite generally, in polaronic systems two parameters are controlling their 
physics: (i)~the strength of electron lattice coupling 
$\alpha = \varepsilon_P/\hbar \bar{\omega}$ ($\bar{\omega}$ denoting some 
averaged phonon frequency and $\varepsilon_P$ the polaron energy level shift) 
and  the adiabaticity ratio $t/\bar{\omega}$ ($t$ denoting the bare electron 
hopping integral defining the electron dispersion 
$\varepsilon_{\bf k} = \frac{t}{z}\sum_{\delta}e^{i{\bf k} \cdot \delta}$, $z$ 
being the coordination number and $\delta$ the lattice vectors linking nearest 
neighbor sites. Let us now consider two limiting cases of such unitary 
transformations in terms of these parameters.

\subsection{Weak coupling adiabatic limit}

In this case we can restrict ourselves to states consisting of electrons 
accompanied by not more than a single phonon at any given time and thus can
limit ourselves to terms in the tranformed Hamiltonian being of quadratic order in 
$V_{\bf q}$. This imposes the condition $[H_{\rm el},S]_- + H_{\rm ph} = 0$ which  
determines the parameters $\pi$ and $\chi$ 
\begin{eqnarray}
\pi_{(j-i,\;l-i)} &=&  [\pi^+_{(j-i,\;l-i)}  - \pi^-_{(j-i,\;l-i)}], \;
\chi_{(j-i,\;l-i)} = [\chi^+_{(j-i,\;l-i)} +\chi^-_{(j-i,\;l-i)}] 
\nonumber \\
\chi^{\pm}_{(j-i,\;l-i)} &=&
{1 \over 2N} \sum_{{\bf k}\,{\bf q}}
{V_{\bf q} \over \varepsilon_{\bf k} - 
\varepsilon_{\bf k-q} \mp \hbar \omega_{\bf q}}
\sqrt{\left( {2M \omega_{\bf q} \over \hbar} \right)}
e^{i{\bf k} \cdot ({\bf R}_j -{\bf R}_i)}
e^{i{\bf q} \cdot ({\bf R}_i -{\bf R}_l)} \nonumber \\
\pi^{\pm}_{(j-i,l-i)} &=&
{1 \over 2iN} \sum_{{\bf k}\,{\bf q}}
{V_{\bf q} \over \varepsilon_{\bf k} - 
\varepsilon_{\bf k-q} \mp \hbar \omega_{\bf q}}
\sqrt{\left({2 \over M \hbar \omega_{\bf q }} \right) }
e^{i{\bf k} \cdot ({\bf R}_j -{\bf R}_i)}
e^{i{\bf q} \cdot ({\bf R}_i -{\bf R}_l)}.
\end{eqnarray} 
The resulting effective Hamiltonian is
\begin{eqnarray}
\tilde H_{\rm BCS} = \sum_{{\bf k}\sigma} (\varepsilon_{\bf k}-\mu) 
c^+_{{\bf k}\sigma}c^{\phantom +}_{{\bf k}\sigma} \nonumber \qquad\qquad \\
+  \sum_{{\bf k}{\bf k'}{\bf q}}|V_{\bf q}|^2
c^+_{{\bf k+q}\uparrow}c^+_{{\bf k'-q}\downarrow}c_{{\bf k'}\downarrow}
c_{{\bf k}\uparrow}{\hbar \omega_{\bf q} \over (\varepsilon_{\bf k} - 
\varepsilon_{\bf k-q})^2 -(\hbar \omega_{\bf q})^2 }
\label{H-BCS}
\end{eqnarray}
which is nothing but the BCS Hamiltonian. The phyiscs which is behind 
this transformation can be seen by looking at the transformed electron operators 
in real space, i.e., 
\begin{equation}
\tilde{c}^+_{{\bf m}\sigma} = c^+_{{\bf m}\sigma} + \sum_{{\bf i}\,{\bf l}}
c^+_{{\bf i}\sigma}[\pi_{(m-i,\;l-i)}P_{\bf l} + 
\chi_{(m-i,\;l-i)}X_{\bf l}].
\end{equation}
\noindent
$\pi_{(m-i,\;l-i)}$ and $\chi_{(m-i,\;l-i)}$ are  slowly varying functions of 
${\bf R}_m - {\bf R}_i$ which extend over large distances of the order of 
$1/\delta k \simeq (2 a /\pi)(v_F/s)$, given the fact that ${\bf k}$ in the 
expressions for $\pi$ and $\chi$ are restricted to a thin region around 
${\bf k_F}$ controlled by an average phonon energy $s\pi/a$ (a denoting the 
lattice constant and $v_F$ the Fermi velocity).  
$\pi_{(m-i,\;l-i)}$ and $\chi_{(m-i,\;l-i)}$  on the contrary are
relatively strongly peaked functions around ${\bf R}_{\bf l} - {\bf R}_{\bf i}=0$ 
since the phonon wave vectors ${\bf q}$ in the expressions for $\pi$ and 
$\chi$ connect any two  points on the Fermi surface and 
thus cover a wide interval, $[0, 2k_F]$.

\begin{figure}[h!t]
\begin{minipage}[c]{7cm}
\includegraphics[width=7cm]{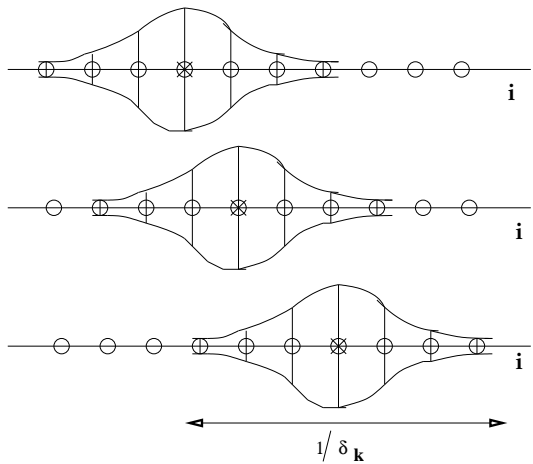}
\end{minipage}
\hspace{0.5cm}
\begin{minipage}[c]{6cm}
\caption{Schematic representation of the transformed electron charge and 
lattice displacements for three successive time steps of the electron moving 
from site to site and characterized by (i)~a delocalization of the electron  over a distance 
$1/\delta_{\bf k}$ controlled by the range of the induced deformation 
centered at a site ${\bf m}$ and indicated by the filled circle, (ii)~an amplitude modulated dynamical lattice deformation.\label{wcBCS}}
\end{minipage}
\end{figure}

This leads to (see fig.~\ref{wcBCS}):

\noindent (i) dynamical deformations  of the lattice around the sites ${\bf l}$ 
(controlled by $\pi_{(m-i,\;l-i)}$),  caused by the presence of 
the electron between sites ${\bf m}$ and ${\bf i}$ some  distance away from 
site  ${\bf l}$ 

\noindent (ii) induce a propagating motion  of the ionic deformations at sites 
${\bf l}$ (controlled by $\chi_{(m-i,\;l-i)}$) through the crystal 
by transferring part of the momentum of the electron to the lattice.

Given the fact that $v_F\gg s$, the dynamical lattice derformation surrounding the 
itinerant electrons can be considered as static. An electron moving in the lattice 
trails with  it a lattice deformation in a kind of tube \cite{Weisskopf-81} 
without radiating off this deformation into the crystal as a whole.  A second 
electron, coming from  the opposite direction, can absorb such a primary 
lattice deformation and it is this which leads to the effective electron-electron 
attraction which is at the heart of the BCS mechanism for Cooper-pair formation.

\subsection{Strong coupling anti-adiabatic limit} \label{antiadiablim}
 
In that case the electron energies $\varepsilon_{\bf k}$ in the denominator 
of the expressions for $\pi$ and $\chi$ can be treated in a perturbative way. 
We can furthermore assume the phonons as being given by dispersion-less Einstein 
modes with frequency $\omega_0$ and a corresponding electron-phonon coupling  
$V_{\bf q} = \lambda \sqrt{\hbar/2M\omega_0}$. To lowest 
order in $\varepsilon_k/\hbar \omega_0$ this yields:
\begin{eqnarray*}
\chi_{(m-i,\;l-i)} = 0\quad\mbox{and}\quad
\pi_{(m-i,\;l-i)} = (i\lambda / \hbar M \omega^2_0)\delta_{im}\delta_{il}.
\end{eqnarray*} 
Treating $\varepsilon_k/\hbar\omega_0$ in a perturbative way reproduces the 
results amounting to the Lang Firsov transformation \cite{Lang-Firsov-63} for small 
polarons, with the operator $S$ reducing to
\begin{eqnarray}
S_{\rm LF}=\sum_{{\bf i}\,\sigma} c^+_{{\bf i}\sigma} 
c^{\phantom +}_{{\bf i}\sigma} 
\pi_{(0,0)} P_{\bf i} =  
-\alpha \sum_{{\bf i}\,\sigma} c^+_{{\bf i}\sigma} 
c^{\phantom +}_{{\bf i}\sigma} 
(a^{\phantom +}_{\bf i} - a^+_{\bf i})
\end{eqnarray}
and with $\alpha = {\lambda \over \sqrt{2 \hbar M \omega^3_0}}$.
The operator $S_{\rm LF}$ transforms the  initial electron on site ${\bf i}$ 
into an electron surrounded by a local lattice deformation
\begin{equation}
\tilde c^+_{{\bf i}\sigma} = c^+_{{\bf i}\sigma}X^+_{\bf i}, \; 
 X^+_{\bf i}|0) = e^{-\alpha (a_{\bf i} -a^+_{\bf i})}|0) = 
\sum_n e^{-\frac{1}{2}\alpha^2}{(\alpha)^n \over \sqrt{n!}} |n)_{\bf i}.
\end{equation}
$H$ transforms correspondingly into an effective polaron Hamiltonian  
\begin{eqnarray}
\tilde H &=& \sum_{{\bf i}\,\sigma} (D - \varepsilon_P) c^+_{{\bf i}\sigma} 
c^{\phantom +}_{{\bf i}\sigma}  - 
t\sum_{{\bf i} \neq {\bf j}, \sigma} (c^+_{{\bf i}\sigma} 
c^{\phantom +}_{{\bf j}\sigma} X^+_{\bf i}X^-_{\bf j} +  H.c.) 
\nonumber \\
&-& 2 \varepsilon_P \sum_{\bf i}c^+_{{\bf i}\uparrow} 
c^+_{{\bf i}\downarrow} 
c^{\phantom +}_{{\bf i}\downarrow}c^{\phantom +}_{{\bf i}\uparrow} + 
\hbar \omega_0 \sum_{\bf i}(a^+_{\bf i} a^{\phantom +}_{\bf i} + \frac{1}{2}),
\end{eqnarray}
whose main physical features have been discussed in my Introductory lecture 
``Introduction to polaron physics: Basic Concepts and models'' in this volume 
and where $\varepsilon_P$ denotes the polaron ionization energy. 

In the limit $2 \varepsilon_P >D$ ($D$ denoting the band half-width),  
bipolaronic  states are stable, i.e., 
the ionization energy of two polarons on different sites ($2 \varepsilon_P$) 
is smaller than of  two electrons on a single site ($4 \varepsilon_P $). 
Approximating  the effective electron hopping term by 
\begin{equation}
tc^+_{{\bf i}\sigma} c^{\phantom +}_{{\bf j}\sigma} X^+_{\bf i}X^-_{\bf j}  \rightarrow 
t^*c^+_{{\bf i}\sigma} c^{\phantom +}_{{\bf j}\sigma},
\end{equation}
with $t^* = te^{-\alpha^2}$  (the Lang-Firsov approximation \cite{Lang-Firsov-63}, 
appropriate for that anti-adiabatic strong coupling case)  we subsequently  
eliminate  all singly occupied sites via a unitary transformation given by 
\begin{equation} 
(S_{\rm BP})_{\alpha \beta} = \sum_{{\bf i}\neq {\bf j}} {(t^*)^2 
(c^+_{{\bf i}\sigma} 
c^{\phantom +}_{{\bf j}\sigma})_{\alpha \beta} \over E_{\alpha} - E_{\beta}}.
\end{equation}
The relevant matrix elements link empty and doubly occupied sites on nearest neighbors with singly occupied ones and $E_{\alpha}, E_{\beta}$ 
denote the energies of single site states of $\tilde H$ in the limit  
$t=0$. The resulting transformed Hamiltonian describes a system of itinerant 
Bipolarons on a lattice \cite{Alexandrov-Ranninger-81} and is given by 
\begin{equation}
H_{\rm BP} = e^{-S_{\rm BP}} \tilde H e^{S_{\rm BP}} = 
-\sum_{\bf i} 2(\varepsilon_P - \mu) (\rho^z-\frac{1}{2}) -
 {(t^*)^2 \over 2 \varepsilon_P} \sum_{<{\bf i} \neq {\bf j}>}(\rho^+_{\bf i} 
\rho^-_{\bf j} + H.c.).
\end{equation}
This corresponds to an effective pseudo-spin-$\frac{1}{2}$ X-Y model in 
an external field, given the fact that  Bipolarons are hard core 
bosons having spin-$\frac{1}{2}$ statistics with
\begin{equation}
\rho^+_{\bf i} = c^+_{{\bf i}\uparrow} c^+_{{\bf i}\downarrow}, \quad 
\rho^-_{\bf i} =  c^{\phantom +}_{{\bf i}\downarrow} 
c^{\phantom +}_{{\bf i}\uparrow}, 
\; \rho^z_{\bf i} = \frac{1}{2} - \rho^+_{\bf i}\rho^-_{\bf i} 
\end{equation}
\begin{equation}
[\rho^-_{\bf i},\rho_{\bf i}^+]_+ = 1, 
\; [\rho_{\bf i}^-,\rho^+_{\bf i}]_- = 
\frac{1}{2} - \rho^+_{\bf i} \rho^-_{\bf i}, \;  
[\rho_{\bf i}^-,\rho^+_{\bf j}]_- = 0 \; ({\bf i} \neq {\bf j}) 
\end{equation}

This is a description analogous to that introduced by Anderson's pseudospin 
representation \cite{Anderson-58} of the effective BCS Hamiltonian, eq.~(\ref{H-BCS}), when restricting Cooper pairing to zero momentum pairs, resulting in:
\begin{equation}
H_{\rm BCS} = -\sum_{{\bf k}} (\varepsilon_{\bf k}- \mu)(\tau^z_{\bf k}-\frac{1}{2}) - 
\sum_{{\bf k}{\bf k'}} 
v({\bf k},{\bf k'}) 
(\tau^+_{\bf k}\tau^-_{\bf k'} + H.c.).
\end{equation}
$H_{\rm BCS}$ has a structure which is formally similar to that of the Hamiltonian
for Bipolarons $H_{\rm BP}$, except that here it is in ${\bf k}$-space rather 
than in real space and with corresponding pseudospin-$\frac{1}{2}$ 
operators   
\begin{equation}
\tau^+_{\bf k} = c^+_{{\bf k}\uparrow} c^+_{{-\bf k} \downarrow}, \quad 
\tau^-_{\bf k} = c^{\phantom +}_{-{\bf k}\downarrow} 
c^{\phantom +}_{{\bf k}\uparrow},\; \tau^z_{\bf k} = \frac{1}{2} - 
\tau^+_{\bf k}\tau^-_{\bf k}
\end{equation}
\begin{equation}
[\tau^-_{\bf k},\tau_{\bf k}^+]_+ = 1, 
\; [\tau^-_{\bf k},\tau^+_{\bf k}]_- =  \tau^z_{\bf k}, \;\;  
[\tau^-_{\bf k},\tau^-_{\bf k'}]_- = 0 \; ({\bf k}\neq {\bf k'}).
\end{equation}

\subsection{Macroscopic phase-locking in superconductivity} 

\begin{figure}[h!t]
\centering{\includegraphics[width= 10cm]{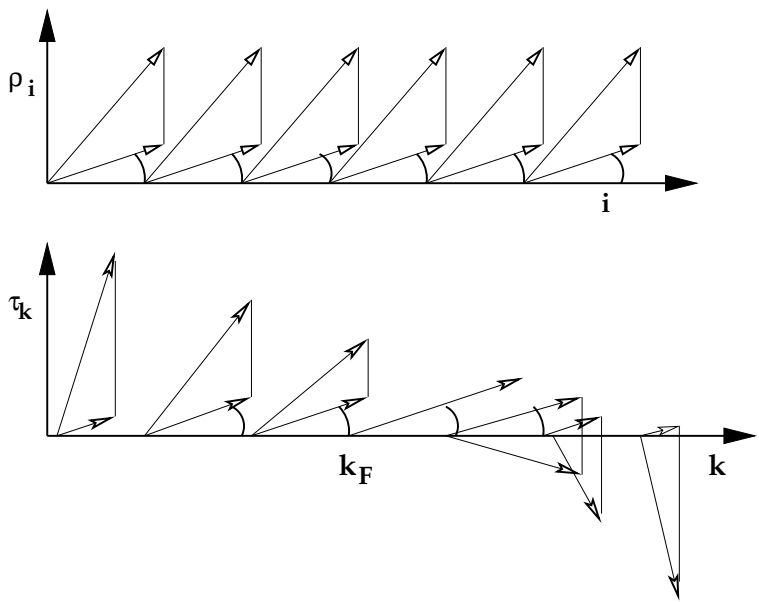}}
\caption{The pseudospin representation of the superconducting ground state, 
manifest in the alignement of the basal plane components of the pseudospin 
vectors indicative of long range phase coherence in the strong  coupling 
Bipolaronic (top figure) and the weak coupling BCS (bottom 
figure) limit.}
\label{figGS}
\end{figure}

The ground states of those two limiting cases of the electron-lattice 
interaction considered above are phase correlated states of the form
\begin{eqnarray}
\Psi^0_{BP} &=& \prod_{\bf i} (u_{\bf i} e^{(\frac{i\phi_{\bf i}}{2})} + 
v_{\bf i} e^{(-\frac{i\phi_{\bf i}}{2})}\rho^+_{\bf i})|0\rangle\\
\label{BP-GS}
\Psi^0_{BCS} &=& \prod_{\bf k}(u_{\bf k}e^{(\frac{i\phi_{\bf k}}{2})} + 
v_{\bf k} e^{(-\frac{i\phi_{\bf k}}{2})})\tau^+_{\bf k}|0\rangle
\label{BCS-GS}
\end{eqnarray}
where  $u_{\bf i}, v_{\bf i} , \phi_{\bf i}$ and 
$u_{\bf k}, v_{\bf k},\phi_{\bf k}$ denote the local amplitudes and phases 
of the Bipolarons and Cooperons respectively and  which, in a macroscopic 
coherent quantum state describing a homogeneous superconducting ground 
state, have to be independent on the sites ${\bf i}$ and respectively the 
wave vectors ${\bf k}$. Superconductivity corresponds to an off-diagonal 
long range order with components of the pseudo-spins which are aligned 
ferro-magnetically in the $x-y$ plane. The modulus of these $x-y$ components 
determine the amplitude of the order parameter. The z-component of the 
pseudo-spins  determine the density of Bipolarons, respectively  of Cooperons.  
The  quantification axes of those pseudo-spins being arbitrary but fixed, 
is chosen here according to the presentation in fig.~\ref{figGS}.
These variational  mean field solutions for the superconducting ground states 
for those two extreme limiting cases, eqs.~(18,19) have distinctively and 
qualitatively different excitation spectra which can be easily visualized by 
inspection of fig.~\ref{figGS}. For the case of Bipolarons, the dominant 
excitations are collective modes 
which  correspond to pseudo-magnons and where, to lowest order, amplitude 
fluctuations play no role. This indicates a temperature driven 
transition from the superconducting into a normal state where Bipolarons 
continue to exist above the superconducting critical temperature $T_c$ up to 
around a certain temperature $T^*$.  Above $T^*$ they 
very rapidly break up into individual electrons and the amplitude of the 
order parameter, even on a short time scale, disappeares. For the case of 
Cooper pairs, the dominant 
excitations which control the breakdown of superconductivity are amplitude 
fluctuations near the Fermi surface which break down the k-space pairing and 
hence cause the vanishing of the amplitude of the order parameter. The 
breakdown of any long range phase correlations is here just a natural consequence 
of the vanishing of the amplitude of the order parameter.
 
The temperature which controls the break-down of these two superconducting  
phases are given respectively by the binding energy of Cooper-pairs (determined 
by  the zero temperature energy gap $\Delta_0$) and the phase 
stiffness of the condensate of the Bipolarons $D_{\phi}$ multiplied by their 
correlation length $\xi$:
\begin{eqnarray}
\Delta_0 \simeq 2 \hbar <\omega_{\bf q}> e^{-{1+\Lambda \over \Lambda}},\; 
\Lambda = N(0){<v^2({\bf k},{\bf k'})> \over  <\omega^2_{\bf q}>},\\
D_{\phi} = \hbar^2(n_{\rm BP}/m_{\rm BP}).\qquad\qquad\qquad
\end{eqnarray}
$<...>$ denote appropriately  chosen averages  over the Fermi surface of the 
phonon frequencies and the electron-lattice couplings. $n_{\rm BP}$ denotes the 
density of the Bipolarons and   
$m_{\rm BP} \simeq (e^{2\alpha^2}/2 \alpha^2 \hbar \omega_0) m_{\rm el}$ ($m_{\rm el}$ 
denoting the bare electron band mass) their exceedingly 
heavy mass \cite{Alexandrov-Ranninger-81}, which makes the chances of ever finding  
such a superconducting state extremely slim. Bipolaronic 
superconductivity definitely is not an explanation for the 
recently discovered high $T_c$ cuprate superconductors \cite{CRF-98} which fall 
in the cross-over regime between BCS and Bipolaronic superconducivity and are 
characterized by Fermionic rather than Bosonic quasi-particles in the normal state.

\section{Resonating bipolarons}\label{resonatingbipolar}

\begin{figure}[h!t]
\includegraphics[width=6.5cm]{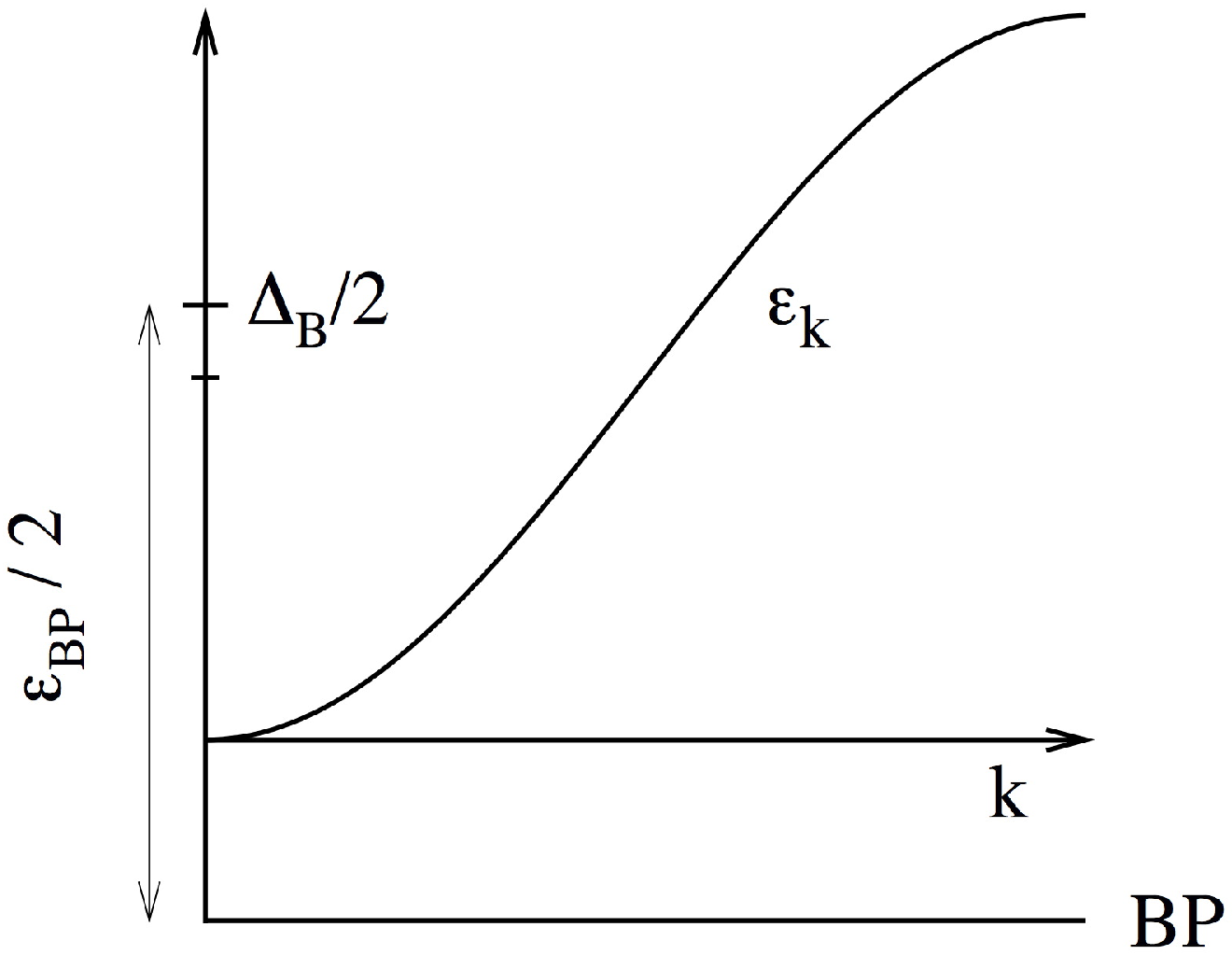}
\includegraphics[width=6.5cm]{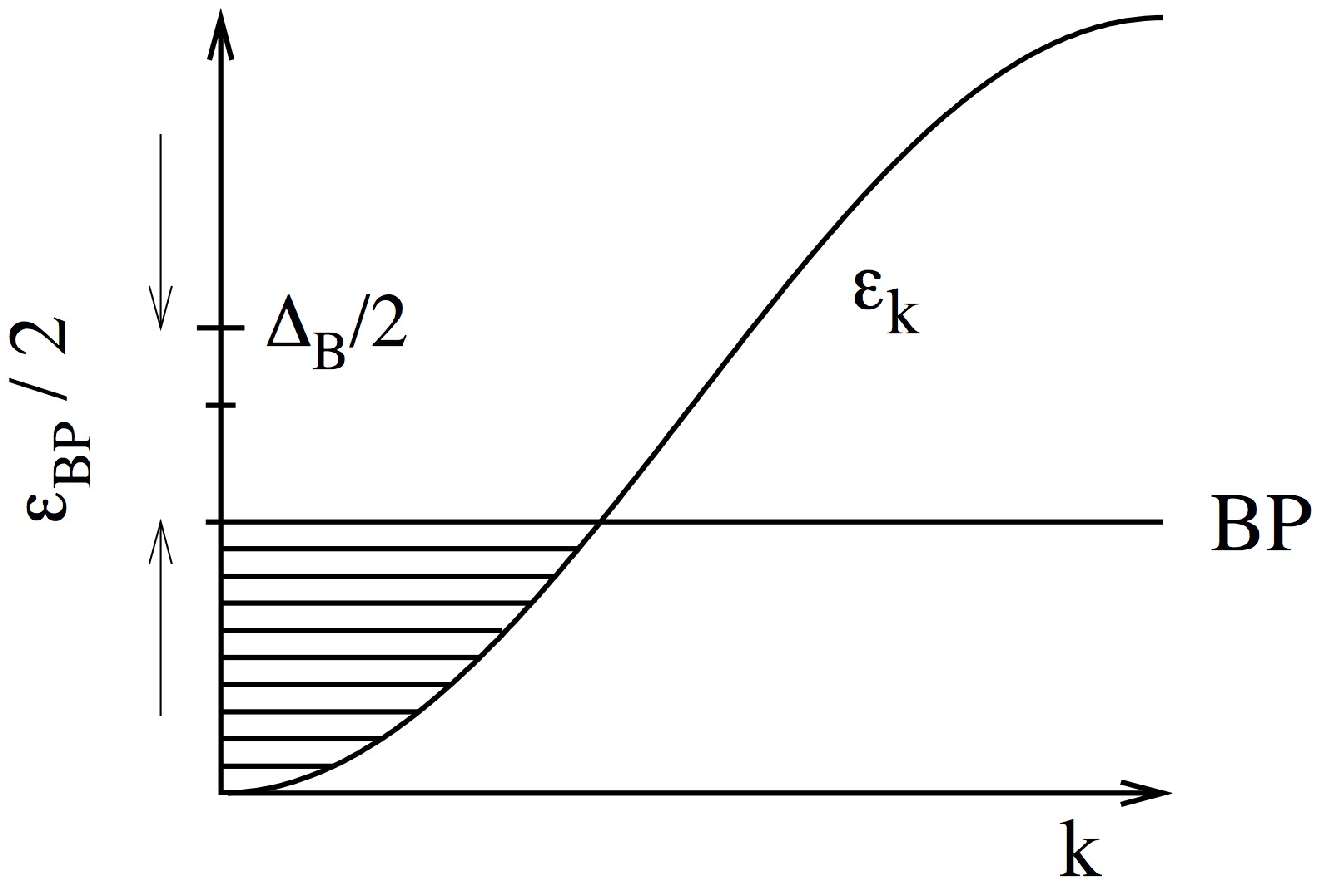}
\hspace*{3.25cm}{\small a)}\hspace*{7cm}{\small b)}
\caption{Schematic plot of the bipolaron level (BP) falling below the band of 
itinerant electrons (a) and inside this band (b).}
\label{band1}
\end{figure}

As we have seen above, as a function of strength of the electron-phonon 
coupling and the adiabaticity ratio, we can obtain two qualitatively distinct 
superconducting phases: 

\noindent (i) a BCS like phase controlled by amplitude fluctuations 
in the weak coupling adiabatic limit,

\noindent (ii) a Bipolaronic Superconductor phase, 
controlled by phase fluctuations of tightly bound electron pairs in the strong 
coupling anti-adiabatic limit when the attractive interaction  between two  
electrons is big enough such as to form true bound states (with a binding energy 
$\varepsilon_{BP}= 4\alpha^2\hbar \omega_0 > 2D$) below the continuum 
of the itinerant electronic states (see fig.~\ref{band1}a). 

\begin{figure}
\begin{minipage}[c]{6.5cm}
\includegraphics[width=6.5cm]{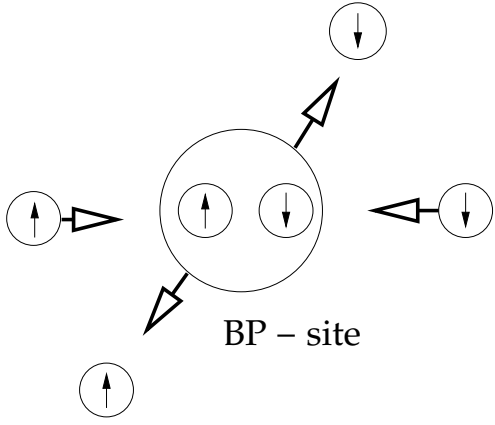}
\end{minipage}
\hspace{0.5cm}
\begin{minipage}[c]{7cm}
\caption{Schematic plot of a resonant process of localized Bipolarons and pairs 
of itinerant electrons.}
\label{Fesh}
\end{minipage}
\end{figure}

Let us now consider the situation when the binding energy $\varepsilon_{\rm BP}$ of 
two electrons, dynamically decoupled from the rest of the remaining 
charge carriers in the system, is less than the band width $2D$. In that case 
we have an overlapp in energy of such localized locally bound pairs and 
the non-interacting itinerant electrons (see fig.~\ref{Fesh}), which  will 
 give rise to resonant scattering between the two species. On the basis of such a 
simple intuitive picture and as a follow-up of the Bipolaronic Superconductor 
scenario \cite{Alexandrov-Ranninger-81}, a phenomenological model was proposed in 
the early eighties---the Boson-Fermion model---devised to capture the main features of 
electron-lattice coupled systems in the cross-over regime between weak and strong  coupling. 
The corresponding Hamiltonian for that is given by 
\cite{Ranninger-Romano-98,Ranninger-Romano-02}:\\

\vbox{
\begin{eqnarray}
H_{\rm BFM} &=& (D-\mu)\sum_{{\bf i},\sigma}n_{{\bf i}\sigma}
-t\sum_{\langle {\bf i}\neq {\bf j}\rangle\sigma}c^+_{{\bf i}\sigma}
c_{{\bf j}\sigma}\qquad \nonumber \\
&&-  (\Delta_{B}-2\mu) \sum_{\bf i} \left( \rho_{\bf i}^z - \frac{1}{2} \right)
+v\sum_{\bf i} [\rho^+_{\bf i}c_{{\bf i}\downarrow}c_{{\bf i}\uparrow}
+\rho_{\bf i}^- c^+_{{\bf i}\uparrow}c^+_{{\bf i}\downarrow}] \nonumber \\
&&-  \hbar \omega_0 \alpha \sum_{\bf i} \left( \rho_{\bf i}^z - \frac{1}{2}
\right) (a_{\bf i}+a_{\bf i}^{+}) +\hbar \omega_0 \sum_{\bf i} 
\left(a^{+}_{\bf i} a_{\bf i}+\frac{1}{2}\right). 
\label{BFM}
\end{eqnarray}}
$\rho^+_{\bf i}, \rho^-_{\bf i},\rho^z_{\bf i}$ denote creation, 
annihilation, and density operators for localized electron pairs coupled 
to local lattice deformations and which, as a result of that coupling, end 
up as localized Bipolarons described by 
$\rho^{+}_{\bf i} \, X_{\bf i}^{+}\equiv c^+_{{\bf i}\uparrow}\,
c^+_{{\bf i}\downarrow}\,e^{-2\alpha(a^{\phantom +}_{\bf i} - a^+_{\bf i})}$. 
We subsequently treat such localized Bipolarons as commuting with the 
itinerant electrons but compensate that transgression by introducing  
phenomenologically a charge exchange term 
of strength $v$ between the two  species. Experimental tests which go along  
with such a scenario are based on measurements of local structural dynamical 
instabilities which show up in pair distribution functions consisting of a 
double peak structure which strongly depends on the time scale of the 
experiments (see the articles by T.~Egami and N.~Saini, this volume). The 
manifestation of a double peak structure of the local lattice environment is 
a consequence of local lattice deformation fluctuation induced by the resonant 
pairing of the itinerant electrons on polaronic sites \cite{Ranninger-Romano-98}. 
This goes hand in hand with 
the evoltion of spectral features in the two-particle properties which go 
from well defined itinerant behavior (characterized by a sharp peak at a given 
frequency for a specific ${\bf k}$ vector) to that of localized features 
(characterized by a broad incoherent background). The resonant behavior which  
lies between these two limiting cases is characterized by a spectral function 
where those two features overlap in frequency.

 Treating the cross-over regime of the electron-lattice coupled systems 
on the basis of this Boson-Fermion model introduces a separation of energy 
scales according to which electron-pair correlations start to build up 
independently of any superconducting long range order. But unlike the strong 
coupling limit, where the state above the superconducting state is one of 
itinerant Bipolarons, here,  the normal state is a system of electrons which 
due to their strong lattice induced attractive interaction show a well pronounced 
pseudogap. This pseudogap only disappears at much higher temperature, when finally 
these local electron-pair  correlations disappear.
The lattice induced origin for pairing giving rise to superconductivity in 
such a scenario can be examined via the isotope effect, not of 
the onset temperature of superconductivity but rather of the onset 
temperature of electron-pair correlations. Although these  
correlations are dynamical in origin and do not show up as a phase transition at 
a given temperature, their onset is very abrupt around a 
certain temperature  $T^*$. One thus can get a  reasonably good indication 
for the isotope effect of electron pairing around  $T^*$ by studying this model 
within a mean field analysis \cite{Ranninger-Romano-04} which determines $T^*$ as 
being exclusively due to
amplitude fluctuations of the electron-pairs. Given this 
situation we can formulate a variational state as the direct product of the 
two components involving localized Bipolarons and pairs of itinerant electrons:
$|\Psi^{\rm F} \rangle \otimes   \prod_{\bf i} |\; l \;\}^{B}_{\bf i}$ with
\begin{eqnarray}
|\Psi^{\rm F} \rangle = \prod_{\bf k} \left[ u_{\bf k} + v_{\bf k} 
c^+_{{\bf k}\uparrow}c^+_{-{\bf k} \downarrow} \right]| 0 \rangle, \quad 
|\; l \;\}^{B}_{\bf i} = \sum_n \left[ u_{l n}(i) +
v_{ln}(i)\,\rho_{\bf i}^+\right] | 0)_{\bf i}|n >_{\bf i} \; .
\end{eqnarray}
$|n >_{\bf i}$ denotes the set of oscillator states on sites ${\bf i}$ 
composed of $n$ phonons.
The exchange coupling term in $H_{\rm BFM}$, eq.~(\ref{BFM}) then  becomes
\begin{eqnarray}
v \rho x + vx\sum_{\bf i} [\rho_{\bf i}^+ + \rho_{\bf i}^-] + 
{v \rho \over 2}\sum_{\bf k}
[ c_{-{\bf k}\downarrow}c_{{\bf k}\uparrow} +
c^+_{{\bf k}\uparrow}c^+_{-{\bf k}\downarrow}]
\end{eqnarray}
with
\begin{equation}
x \; = \; {1 \over N} \sum_{\bf i} \langle c_{{\bf i} \uparrow}^+ c_{{\bf i}
\downarrow}^+ \rangle \, , \quad \rho \; = \; {1 \over N} \sum_{\bf i}
\langle \rho_{\bf i}^+ + \rho_{\bf i}^- \rangle
\end{equation}
denoting the amplitudes of the order parameters of the electron and Bipolaron 
subsystems. The corresponding mean field  Hamiltonian thus reduces to 
\begin{eqnarray}
H_{\rm MFA} & = & H_F +H_B - v \rho x + {\hbar \omega_0 \over 2}
\nonumber \\
H_F & = & (D-\mu)\sum_{{\bf i},\sigma}c^+_{{\bf i}\sigma}c_{{\bf i}\sigma}
-t\sum_{\langle {\bf i}\neq {\bf j}\rangle,\sigma}c^+_{{\bf i}\sigma}
c_{{\bf j}\sigma}
\nonumber\\
&& + {v \rho \over 2}\sum_{\bf i} [ c_{{\bf i}\downarrow}c_{{\bf i}\uparrow}
+c^+_{{\bf i}\uparrow}c^+_{{\bf i}\downarrow}] \nonumber \\
H_B & = & -(\Delta_B-2\mu) \sum_{\bf i} \left( \rho_{\bf i}^z - \frac{1}{2}
\right)+v x \sum_{\bf i} [\rho^+_{\bf i} \; + \rho_{\bf i}^-] \nonumber \\
&-& \hbar \omega_0 \alpha \sum_{\bf i} \left( \rho_{\bf i}^z -\frac{1}{2}
\right) (a_{\bf i}+a_{\bf i}^{+}) +  \hbar \omega_0 \sum_{\bf i} 
(a^{+}_{\bf i} a_{\bf i} + \frac{1}{2}) \; .
\end{eqnarray}
with mean field equations for the order parameters given  by
\begin{eqnarray}
x &=& -{v \rho \over 4 N} \,\sum_{\bf k} \,{1 \over
\tilde{\epsilon}_{\bf k}(\rho)}\,\tanh{\beta\tilde{\epsilon}_{\bf k}(\rho) \over 2} \; ,\nonumber \\
\rho &=& {1 \over Z}\,
\sum_{ln} \, u_{ln} \, v_{ln} \,
\exp\,\left[-\beta E_l(x)\right] \; ,\nonumber \\
n_{tot} &=&  \frac{1}{4}\rho^2 + 2 - {1 \over N}\sum_{{\bf k}}
\left( {\varepsilon_{\bf k} \over \tilde\varepsilon_{\bf k}(\rho)}
\, \tanh{\beta \tilde\varepsilon_{\bf k}(\rho) \over 2}\right) \nonumber \\
&+& \frac{1}{Z} \;\sum_{ln} \left[ (u_{ln})^2
- (v_{ln})^2 \right] \exp\,\left[-\beta E_l(x)\right].
\label{T*}
\end{eqnarray}
$Z=\sum_{l} e^{-\beta E_l(x)}$ denotes  the partition function corresponding 
to the bosonic part of the mean-field Hamiltonian. Such a procedure describes the 
opening of a true gap of size $v\rho$ in the single-particle electron spectrum 
$\tilde\varepsilon_{\bf k}(\rho) = \pm \sqrt{(\varepsilon_{\bf
k}-\mu)^2 + (v \rho)^2/4}$, but under the present physical circumstances should be taken as 
a qualitative description for the real situation where it is a pseudogap 
rather than a true gap which opens up at $T^*$ \cite{RRE-95}. This pseudogap 
plays the role of a precursor 
to a true  superconducting state at some lower temperature which is controlled 
by phase fluctuations \cite{Ranninger-Tripodi-03}. Such a scenario was indeed  
experimentally veryfied in the high $T_c$ cuprate 
superconductors \cite{pseudogap-exp} shortly after this theoretical prediction.

\begin{figure}[h!t]
\centering{\includegraphics[width=8cm]{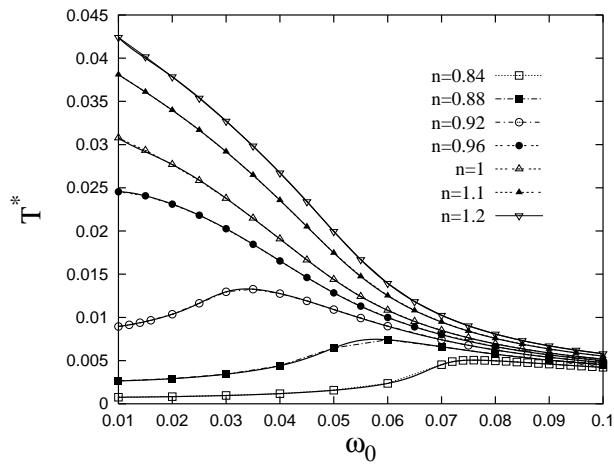}}
\caption{$T^*$ as a function of $\omega_0$ for  $\alpha =2$ and $v=0.25$ 
(after ref.~ \cite{Ranninger-Romano-04}).} 
\label{FigT*}
\end{figure}

Solving the selfconsistent set of eqs.~(\ref{T*}) for $T^*$ (alias the mean field 
critical temperature) for different values of phonon frequencies $\omega_0$ 
and  for a set of concentrations of electrons and localized Bipolarons, we 
illustrate in fig.~\ref{FigT*} its dependence on $\omega_0$. The isotope 
effect in classical weak coupling BCS superconductors  is defined by an exponent 
$\alpha$ which relates the critical temperature for superconductivity to the mass 
of the ions (alternatively to the phonon frequency) via 
$T_c \propto \omega_0^{\alpha}$ and with a value for  $\alpha$ generally around 
$0.5$. In the present scenario, in contrast,  the isotope coefficent for $T^*$ 
itself varies as a function of $\omega_0$. The corresponding isotope exponent then 
has to be defined by the relation
\begin{equation}
\alpha^* = 0.5 \; \ln \left({T^*(i) \over T^*(i+1)}  /
{\omega_0(i) \over \omega_0(i+1)} \right).
\end{equation}
where $\omega_0(i), T^*(i)$ denote a set of different, closely lying together, 
values of phonon frequencies and hence correponding temperatures for the onset of 
pair-corellations. $\alpha^*$  approaches constant positive values in the 
BCS-like limit where Bipolarons are only virtually excited but its value still 
depends sensitively on $\omega_0$. In the case where we have a mixture of 
itinerant electrons and localized Bipolarons (the case for which this model has 
been designed) the isotope exponent deviates  significantly from any BCS 
like behavior. It 
shows  negative and large values which, for the set of parameters chosen in 
fig.~\ref{FigT*}, fall typically between $-0.5$ and $-1$. There are strong 
experimental indications for such large negative values of $\alpha^*$ 
in a number of high $T_c$ cuprates. Experiments  
measuring $T^*$ must be fast enough to capture two electrons remaining
correlated over a finite time, which is typically of the order of the vibrational 
frequency of local modes ($\simeq 10^{-13} - 10^{-15}$ sec), such as to give rise to 
Bipolaron 
formation. Neutron spectroscopy, studying the relaxation rate of crystal field 
excitations and XANES (X-ray absorption near edge spectroscopy) as well as  EXAFS 
are in the right range of time scale and have given results in a number of high 
$T_c$ cuprates \cite{T*exp} such as La$_{2-x}$Sr$_x$CuO$_4$, Ho Ba$_2$Cu$_4$ O$_8$ 
and La$_{1.81}$Ho$_{0.04}$Sr$_{0.15}$CuO$_4$.

\section{Local versus non local phase coherence in the Boson-Fermion model}

In sections \ref{phaseflucsup} and \ref{resonatingbipolar} above we have shown that the amplitude of the order 
parameter 
controls in quite analogous fashions, (i) the onset of superconductivity 
in BCS-like systems and (ii) the onset of electron pairing, but  without any
superconductivity, in systems which are in the cross-over regime between a BCS 
superconductor and a superfluid state of tightly bound electron-pairs. The physics 
leading to the onset of superconductivity in such a
cross-over regime is a subject which presently is far from being 
completely understood. What does  emerge however from preliminary studies is 
that the onset of a condensed superfluid state of Bipolaronic bosons 
has a strong effect on the lattice and generally leads to its stiffening, 
as several examples show: (i) non-interacting itinerant charged bosons coupled 
to acoustic longitudinal phonons manifest an abrupt change in the increase 
of sound velocity as the temperature is reduced to below the condensation 
temperature of the bosons \cite{Alexandrov-Ranninger-89}, (ii) the coupling of 
weakly interacting itinerant bosons to local Einstein modes on a deformable 
lattice results in correlated condensates involving both, the  bosons as well 
as the phonons \cite{Jackeli-Ranninger-01}, (iii) localized Bipolarons coupled 
to a system of itinerant electrons, such as described  by the Boson Fermion model 
scenario, leads to an abrupt change in the decrease of the Debye Waller factor 
as the temperature decreases through 
$T^*$ \cite{Ranninger-Tripodi-99,Ranninger-Romano-02}. All these preliminary 
indications for strong electron-phonon coupling induced macroscopic lattice 
effects (particularly well documented in the study of the high $T_c$ cuprate 
superconductors) can possibly be the reason for: (i)~the onset of dynamically 
ordered vibrations of atoms upon entering the superfluid phase, such as seen in 
Rutherford back scattering experiments \cite{ionchannelling}, (ii)~the abrupt changes 
in the decrease of the kinetic energies of the atomic vibrations \cite{Mook-90}, 
(iii)~changes in the near IR excited Raman scattering structure showing 
an  abrupt increase in the low energy electronic background \cite{Ruani-Ricci-97}, 
(iv)~a sharp 
increase in the intensity of certain Raman active phonon modes suggesting a modification 
of the scattering mechanism \cite{Misochko-99} as well as  
(v)~similar abrupt changes in the sound velocity of certain modes and the Debye 
Waller factors (we refer the reader to the articles by Egami and Saini in this 
volume).

\begin{figure}[h!t]
\begin{minipage}[c]{7cm}
\includegraphics[width=7cm]{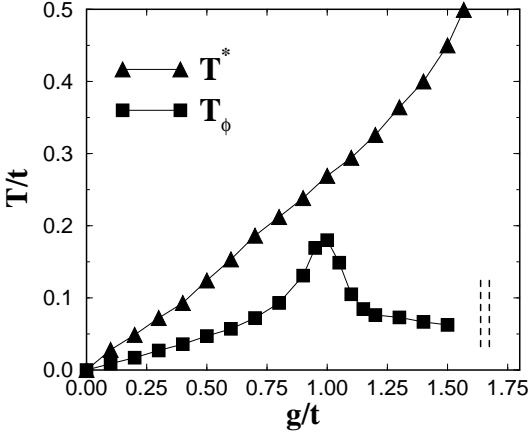}
\end{minipage}
\hfill
\begin{minipage}[c]{5.5cm}
\caption{Evolution of the onset temperatures $T^*$ and $T_\phi$ of local and global phase correlations 
\mbox{$<\rho^+_{\bf i}\tau^-_{\bf i}>$} and  \mbox{$<\rho^+_{{\bf q}=0}\rho^-_{{\bf q}=0}>$} 
as a function of the Boson-Fermion exchange coupling $g$ (after an exact 
diagonalization study \cite{cuoco} on an 8-site ring).}
\label{fig-correl}
\end{minipage}
\end{figure}

Let us conclude this overview of resonating Biplolarons by a discussion of their 
superfluid properties and the possibility of a Superconductor - Bipolaron 
Insulator 
quantum phase transition. Such physics goes beyond the so-called BCS to local pair 
superconductor cross-over which has been widely studied in the 
past \cite{Eagles-86,Leggett-80,Nozieres-Schmitt-Rink-85} on the basis of models 
with an effective attractive interaction between the electrons.  These studies 
invariably show a continuous evolution of a superfluid ground state as the 
strength of this interaction is varied. What is new in the present scenario 
is that the locally fluctuating lattice  deformations---which we believe to 
be the main new feature in this cross-over regime of electron-lattice coupled 
systems---lead to strong local correlations between two types of charge carriers: 
itinerant electrons and localized bound electron-pairs which resonate on a given 
polaronic site. It is this mechanism which not only  
incites pairing of the itinerant electrons and their ultimate superfluid 
features but also limits their long range superfluid phase correlations when 
this exchange coupling goes beyond a certain critical value. In order to reduce 
the complexity of this problem let us now assume the phonons in the clouds 
surrounding the locally 
bound pairs to be  frozen out. This amounts to consider the case of a mixture 
of tightly bound localized electron-pairs (localized Bipolarons where the lattice 
deformations surrounding those charge carriers are static) which are 
exchange-coupled to pairs of itinerant electrons. The corresponding reduced 
Boson-Fermion model is then given by
\begin{eqnarray}
H_{\rm BFM} &=& (D - \mu)\sum_{{\bf i}\sigma} c^+_{{\bf i}\sigma} 
c^{\phantom +}_{{\bf i}\sigma} - 
t\sum_{{<{\bf i},{\bf j}>}\sigma}c^+_{{\bf i}\sigma}c_{{\bf j}\sigma} 
- (\Delta_B - 2\mu) \sum_{\bf i} (\rho^z_{\bf i}-\frac{1}{2}) \nonumber \\
&+& g\sum_{\bf i}(\rho^+_{\bf i}\tau^-_{\bf i} + (\rho^-_{\bf i}\tau^+_{\bf i})
\end{eqnarray}
\noindent
where $g$ denotes some effective exchange coupling constant. 
$\{\rho^+_{\bf i}, \rho^-_{\bf i}, \rho^z_{\bf i}\}$ denote, as before, localized 
tightly bound electron-pairs and 
\begin{equation}
\tau^+_{\bf i} = c^+_{{\bf i}\uparrow} c^+_{{\bf i} \downarrow}, \quad 
\tau^-_{\bf i} = c^{\phantom +}_{{\bf i}\downarrow} 
c^{\phantom +}_{{\bf i}\uparrow},\quad \tau^z_{\bf i} = \frac{1}{2} - 
\tau^+_{\bf i}\tau^-_{\bf i}
\end{equation}
\begin{equation}
[\tau^-_{\bf i},\tau_{\bf i}^+]_+ = 1, 
\; [\tau^-_{\bf i},\tau^+_{\bf i}]_- =  \tau^z_{\bf i}, \;\;  
[\tau^-_{\bf i},\tau^-_{\bf j}]_- = 0 \; ({\bf i}\neq {\bf j})
\end{equation}
pairs of itinerant electrons in real space which are the equivalents of the 
Cooper-pairs $\{\tau_{\bf k}^+,\tau_{\bf k}^-,\tau_{\bf k}^z\}$ in the standard 
BCS scenario discussed in section \ref{antiadiablim}.
In order to illustrate the competition between local and global phase 
coherence let us consider for simplicity's sake and without any loss of 
generality the case $\Delta_B = 0, \mu = 0, n_F = n_B = 1$. In the limit 
where the itinerancy of the electrons can be neglected against the pair-exchange 
coupling, $t << g$, the ground state is given by 
\begin{equation}
\prod_{\bf i}\frac{1}{\sqrt 2} e^{i\phi_{\bf i}}[\rho^+_{\bf i} + 
\tau^+_{\bf i}]|0>.
\label{GSinsul}
\end{equation}
It presents  an insulating  system of locally strongly correlated pairs involving  
the two species of charge carriers, with full phase locking between them but 
arbitrary and  thus uncorrelated local phases $\phi_{\bf i}$ on different sites. 
With decreasing $g/t$ one should expect that the above state goes 
over into a superfluid phase locked state with $\phi_{\bf i} \equiv \phi$ which, to 
within the simplest approximation, could be assumed by the form
\begin{equation}
\prod_{\bf i} (u  + v \rho^+_{\bf i}|0> \otimes (u'  + v' \tau^+_{\bf i}|0>.
\end{equation}
Yet, the strong onsite correlations between the two species of charge carriers, 
necessary to introduce pairing in the Fermion subsystem in the first place, is 
counterproductive to such global superfluid phase correlations as the following 
identity shows
\begin{equation}
\quad\prod_{\bf i}{1 \over \sqrt 2}[\rho^+_{\bf i} + \tau^+_{\bf i}]|0>|0) = 
\frac{1}{2 \pi}
\int_{-\pi}^{+\pi} d\phi_{\bf i} [\cos(\phi_{\bf i}) \tau^+_{\bf i} + 
\sin(\phi_{\bf i}][\sin(\phi_{\bf i}) \rho^+_{\bf i} + 
\cos(\phi_{\bf i})].
\end{equation}
It in fact impies randomizing any potential off-diagonal order which would be 
necessary for establishing a global phase coherent state. The realization of a possible 
superfluid phase in such a scenario must be searched in a compromise between 
local and global phase correlations. Such  
competing behavior can be illustrated upon comparing the evolution of 
the onset temperatures $T^*$ and $T_{\phi}$  controlling the correlators 
$<\rho^+_{\bf i}\tau^-_{\bf i}>$ and $<\rho^+_{{\bf q}=0}\rho^-_{{\bf q}=0}>$ 
as a function of $g/t$. This is shown in fig.~\ref{fig-correl} 
where one can clearly see the initial increase of the superconducting 
correlations together with the increase of local pair correlations as $g/t$ 
increases. Beyond a certain  critical value of $g/t$ however, the superfluid 
correlations break down and the system collapses into an insulating states given 
by expression, eq.~(\ref{GSinsul}).

Let us now formulate this physics in terms of a functional integral representation 
of the Boson-Fermion model, 
which, after integrating out all single electron states,  results in  an effective 
action involving both, amplitude and phase fluctuations of the local 
an itinerant electron pairs \cite{Cuoco-Ranninger-04}. This is done by 
introducing Grassmann variables $\{\bar{\Psi}_i,{\Psi}_i\}$  for the fermionic 
operators in a Nambu spinor representation and via a functional 
integral representation for the quantum spins representing the hard-core bosons 
of the localized electron-pairs on each lattice site. Those hardcore Bosons are parametrized 
by spherical coordinates describing a classical spin of magnitude $s=\frac{1}{2}$
\begin{equation}
\vec {\rho}_{\bf i}= s (\sin\theta_{\bf i} \cos\phi_{\bf i}, \sin\theta_{\bf i} 
\sin \phi_{\bf i}, \cos\theta_{\bf i}).
\end{equation}
The correct quantification of the corresponding quantum pseudo-spin is 
assured by introducing  a topological Wess-Zumino term in the effective 
action which has the form
\begin{equation}
A_{\rm WZ}[\bar{\Psi}_i,{\Psi}_i,\theta_i,\phi_i]=is\int d\tau \sum_i 
(1-\cos\theta_i)~\partial_{\tau} \phi_i.
\end{equation}
\begin{figure}[h!t]
\begin{minipage}[c]{7.5cm}
\includegraphics[width=7.5cm]{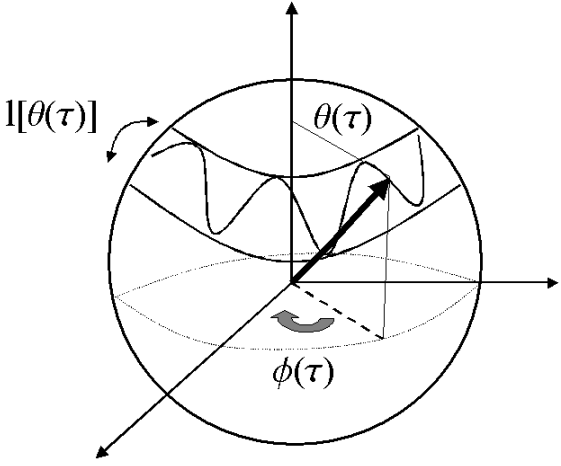}
\end{minipage}
\hfill
\begin{minipage}[c]{5.5cm}
\caption{Illustration of a typical path of the motion of the 
pseudospin on a given site precessing around the $z$-axis and following the 
evolution in time of the phase variable $\phi(\tau))$. $l[\theta(\tau)]$ 
indicates the amplitude of the pseudospin undulations driven  by both, the
fluctuations of the average boson density ($\langle \cos \theta
\rangle$) and of the fluctuations of the pairing amplitude 
($\langle \sin \theta\rangle$) (after ref. \cite{Cuoco-Ranninger-04}).}
\label{FigWZ}
\end{minipage}
\end{figure}

\noindent The resulting effective action is a sum of terms involving the fluctuations 
of the phase $\phi_{\bf i}$ and  the amplitude $sin \theta_{\bf i}$ of the 
pseudospin variable and the interaction involving the two
\begin{eqnarray}
S=\int_0^{\beta} d\tau \left[S_{\phi} + S_{\theta} + S_{\phi-\theta} \right] \,
\label{action}
\end{eqnarray}
Assuming that the amplitude fluctuations are negligible (i.e., for paths of the 
pseudospin moving close to the equator of the sphere in fig.~\ref{FigWZ}, 
which is the case for an density of local pairs around
$n_B \simeq \frac{1}{2}$) we shall
restrict the remaining discussion to a purely phase fluctuation driven 
superfluid state with a corresponding effective action given by 
\begin{eqnarray}
S_{\phi}&=& -E_{J}\sum_{ij}\alpha_{ij} \cos[\phi_i(\tau)-\phi_j(\tau)]
\nonumber \\
&&+\sum_{ij}\frac{1}{8}\frac{\partial \phi_i(\tau)}{\partial \tau} C_{ij}
\frac{\partial \phi_j(\tau)}{\partial \tau}+\sum_{i} \frac{i}{2}
\frac{\partial \phi_i(\tau)}{\partial \tau} q_i(\tau)
\label{Sphi}
\end{eqnarray}
with $\alpha_{ij} = 1$ for $i,j$ denoting nearest neighbor sites and being 
zero otherwise. What is immediately evident from the form of this effective
phase-only part of this action of the Boson Fermion model is its similarity to 
that of Josephson junction arrays with the following correspondences:
\begin{itemize}
\item
$E_J$ controlling the tunneling of the bosons $\Longleftrightarrow$ 
Josephson coupling
\item
$C_{ij}$ controlling the local and non-local density coupling  
$\Longleftrightarrow$ local and non-local capacitance
\item
$q_i$ controlling the amplitude fluctuation induced local charge 
$\Longleftrightarrow$ offset charge.
\end{itemize}
In complete analogy to those widely studied Josephson junction array systems, this 
Boson-Fermion model scenario exhibits a  Superconductor to Insulator transition. 
It is driven by a competition between the pair hopping-induced phase coherence  
and the {\it local charging effect} which is a function of the  local boson 
density (or equivalently pairing-field amplitude) fluctuations. Nevertheless, 
noticeable differences of the action for the Boson-Fermion model and that of the 
Josephson-junction array systems exist:

\noindent
(i) the effective coupling constants now depend in a highly non-trivial 
manner on the original parameters of the underlying Boson-Fermion model hamiltonian which 
gives rise, as we shall see below,  to an intricate dependence of the phase  
diagram on the two parameters, $g/t$ and the boson concentration $n_B$.

\noindent
(ii) the offset charge $q_i =  \mu/g + 1 - cos\theta \equiv  \mu/g + 1 - n_B$ 
is controlled by the topological Wess Zumino term, leading to a time dependence 
of this offset charge  caused by a coupling between the phase and 
the concentration (or alterantively the amplitude)  fluctuations of the 
local pairs. As a consequence the physics of this Boson-Fermion model is 
controlled not only by the parameters $g/t$ and 
$n_{\rm tot} = n_F + n_B$ but also by the ratio $n_F/n_B$ \cite{Cuoco-Ranninger-04}.
Thus, for fixed values of $g/t$ and a total density $n_{\rm tot}$ the parameter which 
controls both $E_J$ and $C_{ij}$ is given by $\tilde g = 2g \sqrt{n_B(n_B-1)}$.
Hence, as $n_B$ approaches either the very dilute or very dense limit 
$n_B= 0,1$, the expected superfluid transition temperature goes to zero, given 
the assumption that the amplitude fluctuations in that case play a minor role. On the 
other hand, by suitably monitoring the total local charge one can obtain a stable 
superfluid state. To extract this type 
of physics one has to resort to a coarse-graining description \cite{Doniach-81} 
with $<e^{i\phi_{\bf i}}>$ taken as an order parameter. The derivation 
of that is standard but rather involved and we refer the reader to refs. 
\cite{Doniach-81,Cuoco-Ranninger-04} for details. Such a procedure  
leads to an effective free energy given by
\begin{equation}
F_\psi=\int_{0}^{\beta}  d\tau  d\tau^{'} \sum_{ij}
\psi_i^{*}(\tau) \left[ \alpha_{ij} \delta(\tau-\tau^{'})-
\chi_{ij}(\tau,\tau^{'}) \right] \psi_j(\tau^{'}),
\end{equation}
 with 
\begin{equation}
\chi_{ij}(\tau,\tau^{'})=\langle e^{i \left[
\phi_i(\tau)-\phi_j(\tau)^{'} \right]} \rangle_0
\end{equation} 
and where $\psi_i^{*}(\tau)$ are auxiliary Hubbard-Stratonovic fields 
conjugated to $<e^{i\phi_{\bf i}}>$. The explicit evaluation of the phase 
correlator \cite{Bruder-92} permits to reformulate 
the effective action, eq.~(\ref{Sphi}), in terms of a Ginzburg Landau 
free energy functional in momentum-frequency representation:
\begin{eqnarray}
F_{\psi}=\frac{1}{\beta L} \sum_{n,k} \psi_k^*(\omega_n)\left[
\alpha_k^{-1} -\chi_{k}(\omega_n)\right] \psi_{k}(\omega_n)\, .
\end{eqnarray}
An expansion in terms of small momentum vectors $k$ and Matsubara
frequencies $\omega_n$ yields
\begin{eqnarray}
F_{\psi}=\frac{1}{\beta L} \sum_{k,n} \left[\frac{2}{z E_J}
-\chi_0 + a k^2+b \omega_n^2+ i \lambda \omega_n +...\right]
|\psi_k(\omega_n)|^2 \nonumber \\
\end{eqnarray}
where $b$, and $\lambda$ are the coefficients of the
expansion of $\chi_{k}(\omega_n)$ around the limit $\omega_n=0$.

The superfluid phase separation line is then determined by  the conditions 
that the coefficient of the quadratic terms in $k$ and $\omega_n$ vanish, i.e.,
\begin{eqnarray}
1-\frac{z E_J}{2} \chi_0(0)=0 \,. \label{boundary}
\end{eqnarray}
\begin{figure}[b]
\centerline{\includegraphics[width=7.5cm]{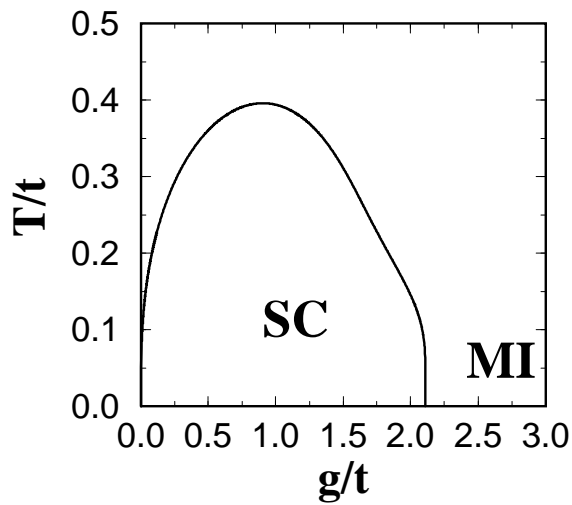}} 
\caption{Phasediagram of the BFM showing as a function of $g/t$ the boundary 
between a Mott insulating (MI) state with homogeneous charge distribution and
a superconducting (SC) phase with long range phase coherence in the representative case of $\Delta_B=0, n_{F\uparrow} = n_{F\downarrow} = 2n_B =1$) (after ref.~\cite{Cuoco-Ranninger-04}).} 
\label{fig4-phasediagPI1}
\end{figure}

The phase diagram which results from this is illustrated  in fig.~\ref{fig4-phasediagPI1}. 
It tracks the gross features of the abrupt beakdown 
of the superfluid phase with increasing $g/t$ as already forshadowed in a 
study based on small clusters \cite{cuoco} which shows the  building up and 
subsequent breaking down of the phase correlations, fig.~\ref{fig-correl}. 
These findings clearly suggest that strong coupling electron-lattice systems in 
the cross-over regime of adiabaticity can truely give rise not only to a 
superfluid phase but also to an insulating Bipolaron state. We stress however that
such an insulating state of Bipolarons would be the result of a non-translational 
symmetry breaking cooperative phenomenon, similar to that of the Mott transition 
in correlated systems. It thus differs significantely from an insulating 
state of disordered non-overlapping localized Bipolarons on a lattice such as could be 
conjectured on the basis of a semi-classical approach and  presented in
 section \ref{resonatingbipolar} in my introductory lecture: ``Introduction to 
polaron physics: concepts and model''.

\section{Conclusion}

Studies devoted to the cross-over from the weak to strong coupling regime and  
from the adiabatic to the non-adiabatic limit generally 
involves the concept of lattice polarons, such as formulated in the paradigme 
of the well studied Holstein model. This model here plays the same role as 
the Hubbard model for systems with strong electron correlations. Here as well 
as there, studying real materials can bring in new and hiherto perhaps unsuspected 
features which might turn out to be of considerable relevance. In the treatise 
developed  in this lecture I have been guided by the well established theoretical 
results of the abrupt change-over from itinerant to almost localized 
polaronic states (so well documented in the various theoretical lectures 
presented here) and conjectured that this scenario would hold in a dense 
polaronic system.  To within a first approximation we can describe such 
physics in terms of an effective model---the Boson-Fermion model---involving  
itinerant non-interacting electrons on an undeformable lattice and 
localized Bipolarons with an exchange coupling between the two. Such a 
picture contains only two relevant parameters: the ratio of the exchange 
coupling to the electron hopping integral and the relative density of 
charge carriers of the two species; the adiabiticity ratio being taken of the order 
unity. Sofar it has not been possible to derive 
this effective Boson-Fermion model scenario from the Holstein Hamiltonian. Yet,
it permits to relatively easyly obtain results on the spectral properties of the 
electrons, the phonons, local dynamical lattice deformations etc., which can be 
tested on real materials and which in fact can and have confirmed such theoretical 
predictions. It might well turn out in the near future that studies of 
real materials which show cross-over features will require some perhaps 
substantial revisions of the Holstein model in the sense that local Martensitic 
like lattice instabilities have to be incorporated in form of some local polaron 
physics and which would involve the interplay of the metal ions and  their ligand 
environements. Such studies are presently being undertaken, but a discussion of 
that is too untimely and should be defered to some later date.

One of the key points I wanted to promulgate in this lecture  was the 
possibility of a Superconductor to  Bipolaronic Insulator phase transition for polaronic 
systems in the strong coupling  regime and in the cross-over between adiabatic and 
anti-adiabatic behavior ($\omega_0/t \simeq 1$)  which is believed to be describable 
within the Boson-Fermion model scenario. The superfluid phase which results in such 
a situation contains features which are common to those encountered in classical 
BCS superconductors (showing the opening of a gap in the single particle electron 
spectrum) and in supefluid systems with collective phase oscillations (showing 
soundwave like excitations in the Cooperon channel) \cite{Domanski-03}. 
We found as an alternative to such a superfluid phase an insulating phase, reminiscent 
of a Mott insulator caused by the predominance of local correlations between 
the Bipolarons and itinerant electron pairs over their long range phase correlations 
between them. This Bipolaron Insulator is quite different from a conceivable 
state of localized Bipolarons randomly distributed on a lattice and which would come 
about by argumenting on the basis of a semi-classical discription  of the 
polaron problem, such as presented in my Introductrory course to polaron physics 
(``Introduction to polaron physics: Basic concepts and Models'') in this volume 
(see in particular the discussion around Fig.~\ref{wcBCS} in section \ref{resonatingbipolar} 
of this contribution). The insulating state discussed here results from a collective phenomenon.

\section{Acknowledgements}
This lecture is dedicated to B. K. Chakraverty. Following  my 
critizisms of his proposed Bipolaronic Insulator phase \cite{Chakraverty-79}, 
his incessant attacks on the competing idea of a
Bipolaronic Superconductor phase \cite{Alexandrov-Ranninger-81} motivated me to
 propose the  scenario of resonating Bipolarons 
and the Boson-Fermion model in the early eighties and which was subsequently 
studied in detail over the past twenty years in Grenoble. I acknowledge the active 
participation of my collaborators on that, and particularly:  M. Cuoco, 
T. Domanski, E. de Mello, T. Kostyrko, M. Robin and  A. Romano.

\end{document}